\documentclass[12pt]{iopart}

\usepackage{graphicx}
\usepackage{color}
\usepackage{bibentry}
\nobibliography*

\usepackage{iopams}  
\begin{document}

\title[`Metal'-like transport in high-resistance, high aspect ratio 2DEGs]{`Metal'-like transport in high-resistance, high aspect ratio two-dimensional electron gases}

\author{Dirk Backes$^1$, Richard Hall$^1$, Michael Pepper$^2$, Harvey Beere$^1$, David Ritchie$^1$ and Vijay Narayan$^1$}

\address{$^1$Cavendish Laboratory, University of Cambridge, J J Thomson Avenue, Cambridge CB3 0HE, UK}
\address{$^2$ Department of Electronic and Electrical Engineering, University College London, Torrington Place, London WC1E 7JE, UK}
\ead{$^*$vn237@cam.ac.uk}

\begin{abstract}
We investigate the striking absence of strong localisation observed in mesoscopic two-dimensional electron gases (2DEGs) (Baenninger \emph{et al} 2008 \emph{Phys. Rev. Lett.} \textbf{100} 1016805, Backes \emph{et al} 2015 \emph{Phys. Rev. B}  \textbf{92} 235427) even when their resistivity $\rho >> h/e^2$. In particular, we try to understand whether this phenomenon originates in quantum many-body effects, or simply percolative transport through a network of electron puddles. To test the latter scenario, we measure the low temperature (low-$T$) transport properties of long and narrow 2DEG devices in which percolation effects should be heavily suppressed in favour of Coulomb blockade. Strikingly we find no indication of Coulomb blockade and that the high-$\rho$, low-$T$ transport is exactly similar to that previously reported in mesoscopic 2DEGs with different geometries. Remarkably, we are able to induce a `metal'-insulator transition (MIT) by applying a perpendicular magnetic field $B$. We present a picture within which these observations fit into the more conventional framework of the 2D MIT.

\end{abstract}


The so-called 2D MIT is the phenomenon wherein $\mathrm{d}\rho / \mathrm{d}T$ of a 2DEG changes from positive to negative at a critical value of carrier concentration $n_\mathrm{s}$, constituting a regime of 'metal'-like transport. The 2D MIT has been widely observed and studied~\cite{Abrahams:2001, Simmons:1998, Hamilton:1999, Huang:2007}, although the definite experiments investigating the quantum corrections which destroy the metallic states at very low T have not been investigated. Thus, the true nature of the ground state, i.e. whether or not it is metallic, remains controversial. Experimentally, it is seen that the critical carrier concentration is not a universal value but rather, the transition occurs when $\rho \approx h/e^2$, the 2DEG being in the metal-like regime for lower $\rho$-values, and in the \textit{strongly localised} or \textit{hopping} regime for higher $\rho$-values. 

The only apparent exception to the above described 2D MIT phenomenology has been observed in 2DEGs with \textit{mesoscopic} dimensions~\cite{Baenninger:2008, Backes:2015, Koushik:2011}. When these were tuned to $\rho >> h/e^2$, nominally deep within the strongly localised regime, the $T$-dependence of $\rho$ was found to be split between metal-like behaviour for $T < 1$~K and localised behaviour for $T > 1$~K. This was subsequently supported by thermopower $S$ measurements~\cite{Narayan:2012, Narayan:2013, Narayan:2014} which showed a metal-like, linear dependence of $S$ on $T$ below $\approx 1$~K.

\begin{figure}
	\centering
	\includegraphics[width=12cm]{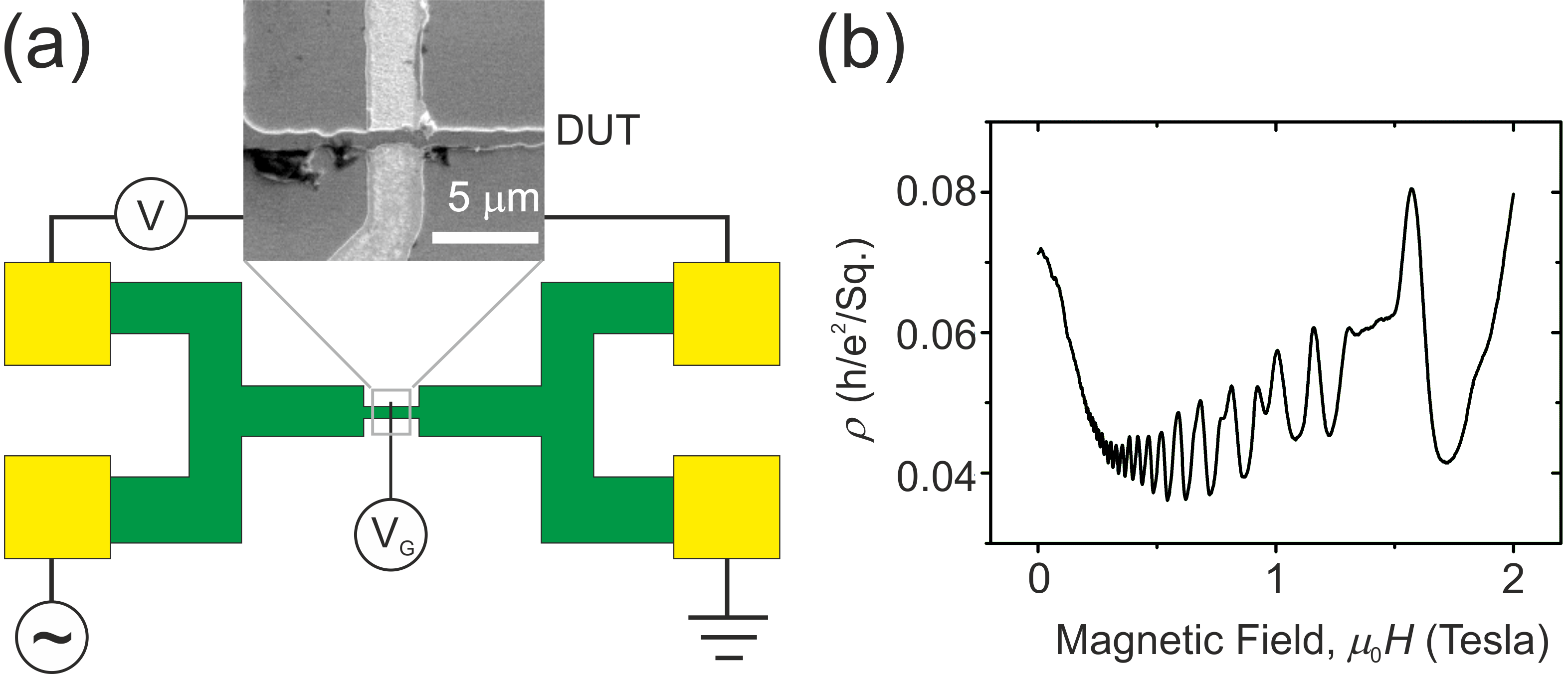}
	\caption{(a) SEM image of the sample geometry. The gate overlaps with the mesa, creating a well-defined depletion zone. (b) Shubnikov-de-Haas oscillations in the longitudinal resistivity upon increasing of a perpendicular magnetic field.}
	\label{fig1}
\end{figure}

There have been contrasting approaches to understand why the strongly localised behaviour is absent in mesoscopic 2DEGs even when $\rho$ is several orders of magnitude larger than $h/e^2$. These range from charge density wave formation~\cite{Baenninger:2008, Koushik:2011}, to scaling behaviour resultant from phase coherent transport~\cite{Backes:2015}, to tunnelling through fluctuations depending on the energy-dependence of wavefunction decay rate~\cite{Backes:2015, Mirlin:2000, Cuevas:2007}, to simply percolative transport through disorder-induced metallic and insulating domains~\cite{Tripathi:2006, Tripathi:2007, Neilson:2011}. In particular, Neilson and Hamilton (NH)~\cite{Neilson:2011} pointed out that the low-$T$ transport through a disordered landscape of electron `puddles' can be entirely dominated by $T$-independent tunnelling and co-tunnelling processes, giving rise to precisely the experimentally observed behaviour. The basic idea behind the NH picture is that in a network of isolated conducting puddles there exist several parallel conducting paths, the least resistive of which will dominate the transport. Importantly, the wide aspect ratio of the mesoscopic devices studied in Ref.~\cite{Baenninger:2008} should facilitate the formation of such low-resistance paths, thus inducing the experimentally observed metal-like behaviour. 

NH further suggested that in order to distinguish between the disorder-based scenario and others, it is necessary to observe the transport in long, narrow 2DEGs. Here, in a puddle-based scenario, the number of low-resistance paths is significantly reduced, increasing the likelihood of Coulomb blockade effects~\cite{Heinzel:2007} in the transport. Coulomb blockade is the phenomenon whereby the transport through nano-scale metallic regions oscillates between zero (blockaded) and non-zero values as a function of the chemical potential $\mu$. This arises due to the electrostatics-driven discretisation of energy levels in the nano-sized metallic domain. Thus the goal of this manuscript is to resolve this very important matter by studying 2DEGs with a length $L$ to width $W$ ratio of $>$ 1, and as small as possible $W$. We note that previous scanning-probe experiments indicate the typical disorder lengthscale in high-mobility GaAs-based 2DEGs to be $\approx 0.5\,\mu$m \cite{Chakraborty:2004}. Motivated by this, the devices in this manuscript have $W$ comparable to this lengthscale. 

\begin{figure}
	\centering
	\includegraphics[width=14cm]{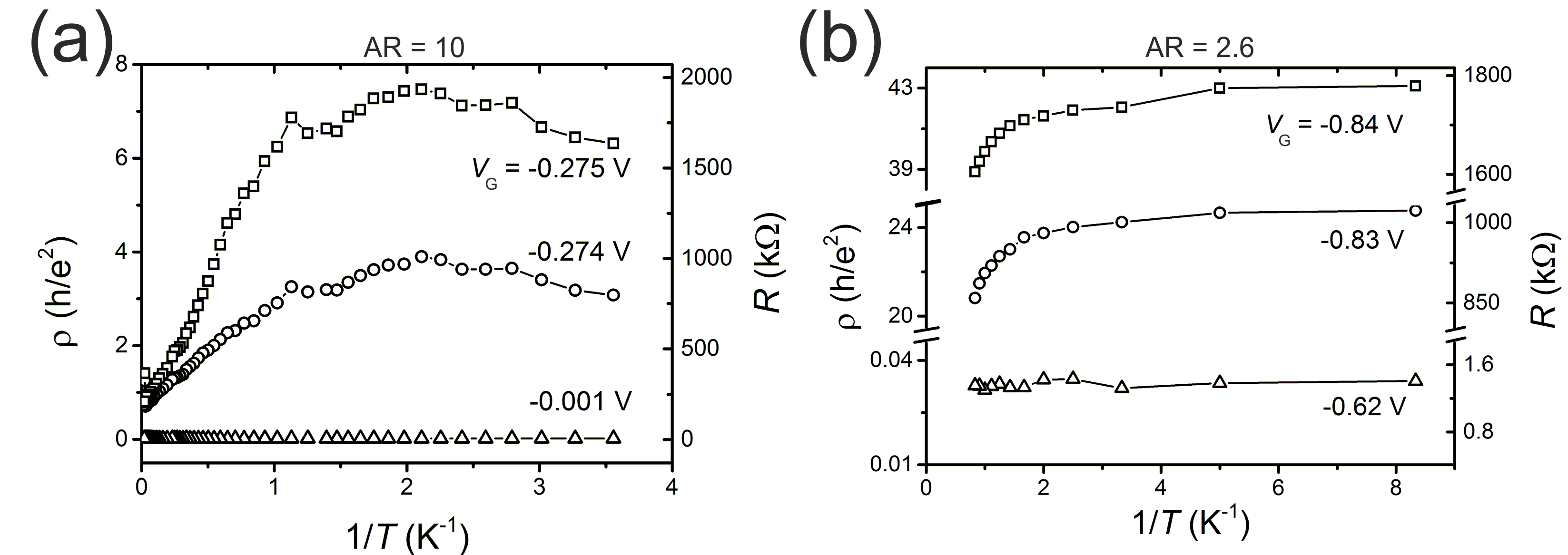}
	\caption{$\rho$ vs. $T$ at different $V_\mathrm{G}$. $W \times L = $ (a) 1~$\mu$m~$\times$~10~$\mu$m (b) 3~$\mu$m~$\times$~8~$\mu$m. The data was taken in a four-terminal measurement setup.}
	\label{fig2}
\end{figure}

We employed GaAs/AlGaAs heterostructures for our experiments in which the 2DEG formed $\approx$~300~nm below the surface. The samples are $\delta$-doped with dopants being situated 40\,nm above the 2DEG. At $T = 4$~K the 2DEG had a mobility $\mu~=~212~\mathrm{m^2 V^{-1} s^{-1}}$ at $n_s=\mathrm{2.3\times10^{15}~m^{-2}}$, as determined from macroscopic Hall bar samples fabricated from the same wafer as the devices investigated. Devices were fabricated using a subtractive process based on wet chemical etching. Photoresist was used as etch mask to create a narrow mesa, and the minimum feature size after etching was $\approx 1~\mu$m, thus setting the smallest achievable $W$. Ohmic contacts were fabricated using a lift-off process based on photolithography and thermal evaporation of AuGeNi alloy. Subsequently, metallic gates were fabricated using a similar lift-off process and thermal evaporation of 15~nm Ti and 120~nm of Au. As shown in figure~\ref{fig1}(a), the high aspect ratio ($L/W$) devices are defined by the overlap of the top-gate and mesa. The gates allowed the application of electrostatic fields, creating a zone of reduced $n_s$ when a negative voltage $V_{\mathrm{G}}$ was applied. The devices were measured either in a He-3 cryostat or dilution refrigerator, each of which has a superconducting magnet. Both four-terminal and two-terminal measurement setups were used to measure the resistance $R$ and conductance $G$ of the 2DEGs, respectively. Figure~\ref{fig1}(a) shows $R$ measured in a four-terminal constant current setup. We used a low excitation current $I_{\mathrm{ex}} = 100$~pA at a frequency $f = 7$~Hz in our experiments. The range of $V_{\mathrm{}}$ was restricted such that $R$ never exceeded 2~M$\Omega$. Figure~\ref{fig1}(b) shows the zero-$V_{\mathrm{}}$ magnetoresistance of the device shown in figure~\ref{fig1}(a). The clear appearance of Shubnikov-de Haas oscillations indicates the existence of a high-mobility 2DEG even in the narrow channel.

Figure~\ref{fig2} shows $\rho$ as a function of $1/T$ at different $V_{\mathrm{G}}$ for two 2DEGs with aspect ratio 10 and 2.6, respectively. The right axis shows the corresponding $R$-values. For low $|V_{\mathrm{G}}|$ the 2DEG is clearly in the metal-like regime with $\rho << h/e^2$ and almost completely independent of $T$. However, for larger $|V_{\mathrm{G}}|$ when $\rho > h/e^2$, $\rho(T)$ is strikingly consistent with metal-like behaviour below $\approx~$0.5~K and with hopping behaviour above $\approx~1$~K, in strong agreement with previous findings~\cite{Baenninger:2008, Backes:2015, Koushik:2011}. At no value of $V_{\mathrm{G}}$ was the low-$T$ metal-like behaviour observed to give way to strongly localised behaviour and this is fundamentally inconsistent with Coulomb blockade where hopping-like transport characteristics are expected. This puts stringent limits on the relevance of the NH picture to our experimental results since it is extremely unlikely that if, indeed, the 2DEG is fragmented into conducting and non-conducting islands, Coulomb blockade is still averted in these narrow 2DEGs. Therefore, this is the first piece of evidence supporting the notion that the 2DEG remains homogeneous at $\rho > h/e^2$.

\begin{figure}
	\centering
	\includegraphics[width=12cm]{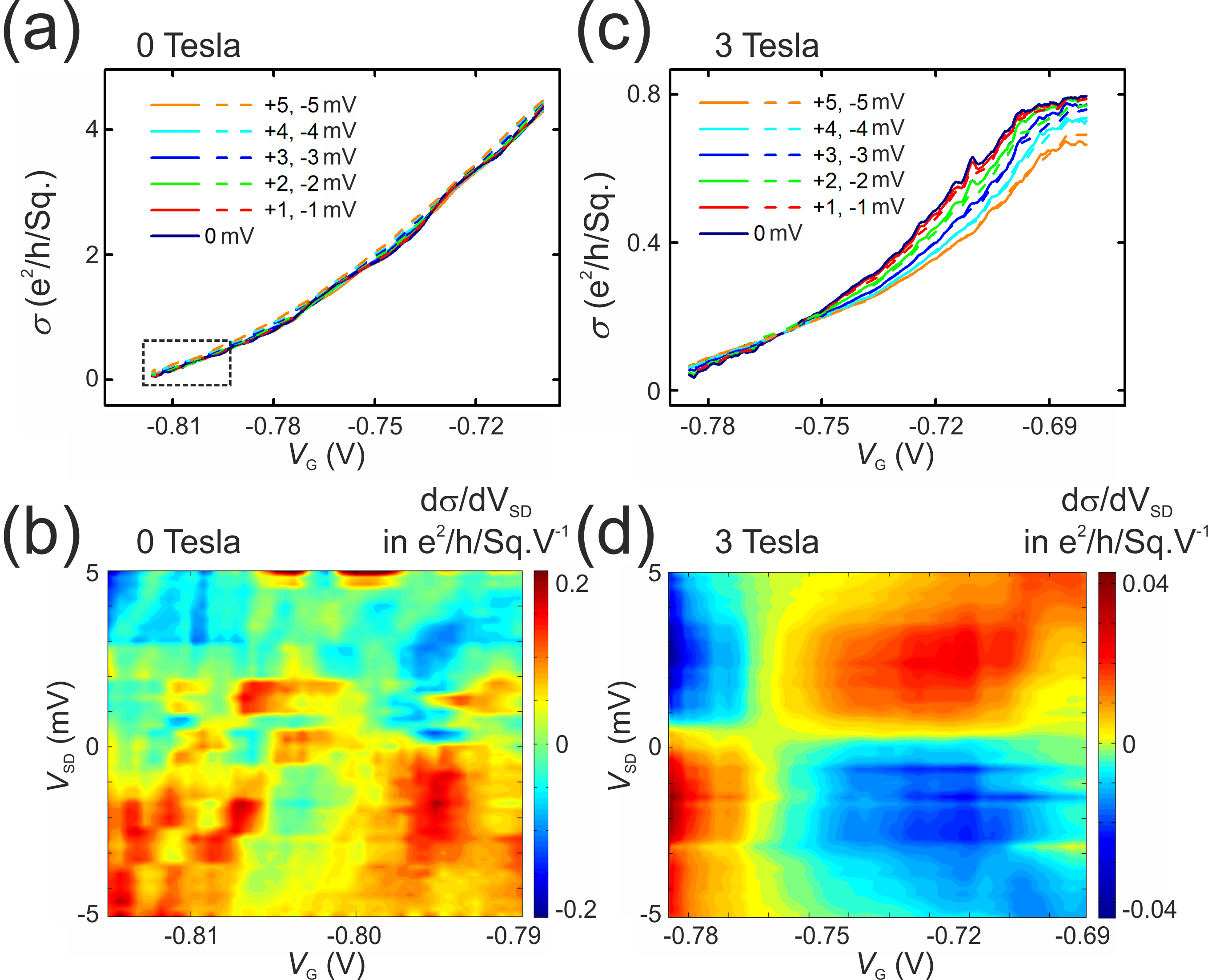}
	\caption{(a) + (c) $\sigma$ as  a function of  $V_{\mathrm{G}}$ for different $V_{\mathrm{SD}}$ at magnetic fields of (a) 0\,T and (c) 3\,T. (b) + (d) Map of $\mathrm{d}\sigma/\mathrm{d}V_{\mathrm{SD}}$ as a function of the gate voltage $V_{\mathrm{G}}$ and source drain bias $V_{\mathrm{SD}}$ at magnetic fields of (b) 0\,T and (d) 3\,T. The range of $V_\mathrm{{G}}$ in (b) is indicated by a dashed box in (a). Data is shown for the aspect ratio 2.6 device.}
	\label{fig3}
\end{figure}

We now investigate whether the dependence of $\rho$ on $V_{\mathrm{G}}$ contains signatures of Coulomb blockade. As briefly described earlier, Coulomb blockade is observed in the electronic transport through a microscopic conducting island (or quantum dot) connected to conducting leads via tunnel barriers. A unique signature of Coulomb blockade is resonant tunneling when the energy levels in the quantum dot match those in the adjacent leads; at resonance $G$ through the quantum dot is finite, but zero otherwise. The resonance condition can be achieved by using an external $V_{\mathrm{G}}$ which shifts the energy levels of the quantum dot, or by applying a bias voltage $V_{\mathrm{SD}}$ between the source and drain leads to directly tune $\mu$ in these. Figure~\ref{fig3}(a) shows $\sigma$ (measured in a two-terminal setup) as a functions of $V_{\mathrm{G}}$ at different $V_\mathrm{{SD}}$ for the device with aspect ratio 2.6. We note that the conductivity $\sigma \sim e^2/h/Sq.$ (where Sq. denotes square), which is roughly where Coulomb blockade can be expected, but $\sigma$ is clearly non-zero. A closer look reveals some modulation of $\sigma$ with $V_{\mathrm{G}}$ when $V_\mathrm{SD}=0$~V. Such modulations have been observed before \cite{Baenninger:2008, Koushik:2011} and are perfectly reproducible between repeated back and forth sweeps. Importantly, however, a finite $V_{\mathrm{SD}}$ has little effect on $\sigma$ and this, again, is inconsistent with Coulomb blockade. 

Figure~\ref{fig3}(b) shows a 2D plot of the derivative $\mathrm{d\sigma / d}V_{\mathrm{SD}}$ of the data in figure~\ref{fig3}(a), specifically focusing on $\sigma \lesssim 0.5\,e^2/h$, the region of conductivity in which the metal-like temperature dependence is observed (see figure~\ref{fig2}b). This corresponds to gate voltages of $-0.815\,V < V_G < -0.79\,V$, indicated by a box in figure~\ref{fig3}(a). We note the slight shift in the pinch-off voltage between figure~\ref{fig2}(b) and figure~\ref{fig3}(a), associated with the additional lead resistance that appears in the two-terminal compared to the four-terminal setup. Figure~\ref{fig3}(b) shows little resemblance to the diamond-shaped pattern that appears in Coulomb blockaded systems~\cite{Heinzel:2007}.

Figures~\ref{fig3}(c) and \ref{fig3}(d) show $\sigma(V_{\mathrm{G}}, V_{\mathrm{SD}})$ in the presence of a perpendicular magnetic field $B = 3$~T. This should confine the electron trajectories to an even narrower region of the sample due to edge state formation and thus make any Coulomb blockade-related effect more visible. We note that the strong $B$-field serves to suppress $\sigma$ by approximately one order of magnitude. Strikingly, however, the effect of $V_{\mathrm{SD}}$ is much more dramatic than in the $B = 0$~T case inducing a qualitative change in the dependence on $V_{\mathrm{SD}}$ at $V\mathrm{_G} \approx -0.76$~V: $\sigma$ decreases (increases) for increasing $|V_{\mathrm{SD}}|$ for $V_{\mathrm{G}}<-0.76$~V ($V_{\mathrm{G}}>-0.76$). This behaviour becomes better visible in figure \ref{fig3}(d), where a map of $\mathrm{d\sigma / d}V_{\mathrm{SD}}$ is shown. A pronounced, symmetric pattern becomes visible, having no resemblance at all with Coulomb diamonds~\cite{Heinzel:2007}. And thus we find that the transport in the measured 2DEGs at $\sigma < 0.5\,e^2/h$ is not consistent with Coulomb blockade which, in turn, suggests that either the puddle size in our sample is significantly smaller than  experimentally observed \cite{Chakraborty:2004} or the 2DEGs are not decomposed into metallic and non-metallic domains. This then leads to the remarkable conclusion that despite the 2DEGs being a continuous Fermi sea between the measurement leads, they have $\rho >> (h/e^2)$/Sq.!

Thus there are two questions at this stage: (1) why is $\rho$ so large if the 2DEGs are, indeed, continuous Fermi seas; and (2) do we understand the qualitative shift in the dependence on $V_{\mathrm{SD}}$ in the presence of a $B$-field. In \cite{Backes:2015} where we studied the geometry dependent transport of similar samples, it was shown that the increased $\rho$ could be understood by considering the finite localisation length $\xi$ of the conducting electrons. In particular, if the system size is comparable to $\xi$, then while $\sigma$ will have a metal-like $T$-dependence (since the wavefunction extends across the system), the finite envelope of $\xi$ will visibly curb the ability of electrons to conduct. This same physical picture underlies the crossover from localised to metal-like behaviour in the $T$-domain (similar to figure~\ref{fig2}), where above a certain $T$ the hopping transport exceeds that due to states `extended' across the system dimensions.

To answer the second question, we first ask how $V_{\mathrm{SD}}$ affects the 2DEG which has a continuous energy spectrum rather than the discrete one of a quantum dot/puddle. In \cite{Hamilton:1999} it was argued that a finite $V_{\mathrm{SD}}$ increases the temperature of the 2DEG and the authors observed a similar effect as in figure~\ref{fig3}(c), albeit in macroscopic 2D hole gases. Heating is thus proportional to $\left| V_{\mathrm{SD}}\right|$, and an increase (decrease) in $G$ with $V_{\mathrm{SD}}$ marks an insulating (metal-like) phase, as for $V_{\mathrm{G}}~<~(>)~-0.76$~V. We independently confirmed this assessment by directly heating a 2DEG with aspect ratio ($\approx~$1), leading to the same qualitative dependence of gate sweeps on temperature as on $V_{\mathrm{SD}}$. Thus, our data indicates a recovery of the MIT in the presence of a $B$-field. From an analysis of the Shubnikov-de Haas periodicity (see figure~\ref{fig1}(b), we ascertain that at $V_{\mathrm{G}}~=~-0.76$~V, $n_{\mathrm{s}}~\approx~4.8~\times~10^{14}~\mathrm{m}^2$. This is slightly less than the $7.3~\times~10^{14}~\mathrm{m}^2$ found in \cite{Hamilton:1999} and in good agreement with the $5.1~\times~10^{14}~\mathrm{m}^2$ found in \cite{Simmons:1998}, which were both obtained for macroscopic samples. However, this is a unique instance in which the transition occurs at a much higher $\rho$ than the traditionally observed $\sim~h/e^2$. 

Regardless, it is important to note that the phase diagram of a mesoscopic sample in the presence of a $B$-field seems to approach the well-known behaviour of macroscopic samples, in that a 2D MIT \textit{is} observed. This is obviously not a true quantum phase transition, but simply results from the $B$-induced shrinkage of $\xi$~\cite{Shklovskii:1983, Nguen:1984} which, in turn, renders conduction only possible via hopping. But this has the important implication that the electronic states, dominating the transport properties on mesoscopic lengthscales at very low $T$, have a finite $\xi$ at $B = 0$~T. As the size of the 2DEG is made larger, states with $\xi$ comparable to the system size are less likely to be occupied and hopping conduction becomes prominent at correspondingly lower $\rho$, ultimately converging to the familiar phase diagram observed in macroscopic 2DEGs.

Summarising, both the observation of a MIT and the absence of Coulomb blockade in long and narrow 2DEGs contradicts the validity of the puddle formation theory~\cite{Neilson:2011}. On the contrary, the restoration of signatures of macroscopic behaviour, i.e. the 2D MIT, upon the application of a magnetic field are understandable within scaling theory of phase coherent transport~\cite{Backes:2015}.

We acknowledge funding from the Leverhulme Trust, UK and the Engineering and Physical Sciences Research Council (EPSRC), UK. UK. DB and VN acknowledge useful discussions with Alex Hamilton. Supporting data for this paper is available
at the DSpace@Cambridge data repository. (http://www.repository.cam.ac.uk/handle/1B10/252620)

\section*{References}



\bibliographystyle{iopart-num} 
\bibliography{references}

\providecommand{\newblock}{}
\begin{thebibliography}{10}
\expandafter\ifx\csname url\endcsname\relax
  \def\url#1{{\tt #1}}\fi
\expandafter\ifx\csname urlprefix\endcsname\relax\def\urlprefix{URL }\fi
\providecommand{\eprint}[2][]{\url{#2}}

\bibitem{Baenninger:2008}
Baenninger M, Ghosh A, Pepper M, Beere H~E, Farrer I and Ritchie D~A 2008 {\em
  Phys. Rev. Lett.\/} {\bf 100} 016805

\bibitem{Backes:2015}
Backes D, Hall R, Pepper M, Beere H, Ritchie D and Narayan V 2015 {\em Phys.
  Rev. B\/} {\bf 92} 235427

\bibitem{Abrahams:2001}
Abrahams E, Kravchenko S~V and Sarachik M~P 2001 {\em Rev. Mod. Phys.\/} {\bf
  73} 251

\bibitem{Simmons:1998}
Simmons M~Y, Hamilton A~R, M~Pepper~and E~H~L, Rose P~D, Ritchie D~A, Savchenko
  A~K and Griffiths T~G 1998 {\em Phys. Rev. Lett.\/} {\bf 80} 1292

\bibitem{Hamilton:1999}
Hamilton A~R, Simmons M~Y, Pepper M, Linfield E~H, Rose P~D and Ritchie D~A
  1999 {\em Phys. Rev. Lett.\/} {\bf 82} 1542

\bibitem{Huang:2007}
Huang J, Xia J~S, Tsui D~C, Pfeiffer L~N,  and West K 2007 {\em Phys. Rev.
  Lett.\/} {\bf 98} 226801

\bibitem{Koushik:2011}
Koushik R, Baenninger M, Narayan V, Mukerjee S, Pepper M, Farrer I, Ritchie D~A
  and Ghosh A 2011 {\em Phys. Rev. B\/} {\bf 83} 085302

\bibitem{Narayan:2012}
Narayan V, Pepper M, Griffiths J, Beere H, Sfigakis F, Jones G, Ritchie D and
  Ghosh A 2012 {\em Phys. Rev. B\/} {\bf 86} 125406

\bibitem{Narayan:2013}
Narayan V, Pepper M, Griffiths J, Beere H, Sfigakis F, Jones G, Ritchie D and
  Ghosh A 2013 {\em J. Low Temp. Phys.\/} {\bf 171} 626

\bibitem{Narayan:2014}
Narayan V, Kogan E, Ford C, Pepper M, Kaveh M, Griffiths J, Jones G, Beere H
  and Ritchie D 2014 {\em New J. Phys.\/} {\bf 16} 085009

\bibitem{Mirlin:2000}
Mirlin A~D 2000 {\em Phys. Rep.\/} {\bf 326} 259

\bibitem{Cuevas:2007}
Cuevas E and Kravtsov V~E 2007 {\em Phys. Rev. B\/} {\bf 76} 235119

\bibitem{Tripathi:2006}
Tripathi V and Kennett M~P 2006 {\em Phys. Rev. B\/} {\bf 74} 195334

\bibitem{Tripathi:2007}
Tripathi V and Kennett M~P 2007 {\em Phys. Rev. B\/} {\bf 76} 115321

\bibitem{Neilson:2011}
Neilson D and Hamilton A~R 2011 {\em Phys. Rev. B\/} {\bf 84} 129901

\bibitem{Heinzel:2007}
Heinzel T 2007 {\em Mesoscopic Electronics in Solid State Nanostructures\/}
  (Weinheim: Wiley-VCH)

\bibitem{Chakraborty:2004}
Chakraborty S, Maasilta I~J, Tessmer S~H and Melloch M~R 2004 {\em Phys. Rev.
  B\/} {\bf 69} 073308

\bibitem{Shklovskii:1983}
Shklovskii B~I 1983 {\em Sov. Phys. Semicond.\/} {\bf 17} 1311

\bibitem{Nguen:1984}
Nguen V~L 1984 {\em Sov. Phys. Semicond.\/} {\bf 18} 207

\end{thebibliography}




\end{document}